\documentclass[12pt]{article}
\usepackage{amsmath}
\usepackage{amsfonts}
\usepackage{epsfig}

\begin{document}

\title{State preparation based on Grover's algorithm in the presence of global
information about the state}

\author{
Andrei N. Soklakov and R\"udiger Schack\\
\\
{\it Department of Mathematics, Royal Holloway,
       University of London,}\\
{\it Egham, Surrey TW20 0EX, United Kingdom}}

\date{3 October 2004}
\maketitle

\begin{abstract}
  In a previous paper \cite{SoklakovSchack} we described a quantum
  algorithm to prepare an arbitrary state of a quantum register with
  arbitrary fidelity. Here we present an alternative algorithm which
  uses a small number of quantum oracles encoding the most significant bits of
  the absolute value of the complex amplitudes, and a small number
  of oracles encoding the most significant bits of the phases. The 
  algorithm given here is considerably simpler than the one described in
  \cite{SoklakovSchack}, on the assumption that a sufficient amount of
  knowledge about the distribution of the absolute values of the complex
  amplitudes is available.
\end{abstract}

\section{Overview}

The first step of many quantum computer algorithms is the preparation of a
quantum register in a simple initial state, e.g., the equal superposition of
all computational basis states. Some applications of quantum computers, such
as the simulation of a physical system \cite{Lloyd1996,Zalka1998,Somaroo1999},
require the initial preparation of more general states. Here we
consider the state preparation problem in the case that the Hilbert-space
dimension of the quantum register is so large that listing the complex
coefficients of the state is impractical.

In a previous publication \cite{SoklakovSchack} we have shown how to use
elements of Grover's algorithm \cite{Grover1997} to prepare a register of
$\log_2N$ qubits, with arbitrary fidelity, in an approximation to the state
\begin{equation}  \label{eq:Psi}
|\Psi\rangle = \sum_{x=0}^{N-1} \sqrt{p(x)} \, e^{2\pi i \phi(x)} |x\rangle
\end{equation}
for any probabilities $p(x)$ and phases $\phi(x)$. We assume
that $N$ is an integer power of 2.  Here and throughout the paper, $|0\rangle,
|1\rangle,\ldots$ denote computational basis states.

We will now describe an alternative algorithm to achieve the same goal, i.e.,
to prepare the quantum register in a state $|\tilde\Psi\rangle$ such that the
fidelity, $|\langle\tilde\Psi|\Psi\rangle|$, is close to 1.  In the
description of the algorithm below, we introduce three positive integer
parameters, $T$, $T'$, and $a$. We will indicate how to choose these
parameters, and  derive a lower bound on the fidelity
$|\langle\tilde\Psi|\Psi\rangle|$ in terms of them.
We will also show how the required computational resources scale with these
parameters. This will put us in a position to compare the two versions of the 
algorithm. See Ref.~\cite{SoklakovSchack} for a comparison with the
state-preparation algorithms by Kaye and Mosca \cite{Kaye2001} and Grover and
Rudolph \cite{Grover-0208}.

The functions $p(x)$ and $\phi(x)$ are assumed to be given in the form of
classical algorithms. The function $p(x)$ is used to construct
a set of quantum {\em oracles\/} as follows.  Let $T$ be a positive integer. 
For $k=1,\ldots,T$, we define
\begin{equation} \label{eq:ok2}
O_k(x)= c_k(x) \;,
\end{equation}
where $c_k(x)\in\{0,1\}$ are the coefficients in the binary expansion
\begin{equation}
\sqrt{\eta N p(x)} = \sum_{k=1}^\infty c_k(x) 2^{-k} \;,
\end{equation}
and where $\eta$ is a positive real number, $\eta<1$, such that 
\begin{equation}    \label{eq:etaBound}
p(x)\le{1\over\eta N} \;\;\mbox{ for all }\; x \;.
\end{equation}
We extend this definition beyond the domain of the function $p$ by setting 
$O_k(x)=0$ for $x\ge N$. 
The quantum oracles are unitary operators defined by
\begin{equation} \label{eq:okHat}
\hat O_k|x\rangle = (-1)^{O_k(x)}|x\rangle \;,
\end{equation}
which can be realized as quantum gate sequences using the classical algorithm
to compute the probabilities $p(x)$. For each oracle, we define the number of
solutions
\begin{equation}
N_k=\sum_x{O_k(x)}\,.
\end{equation}

Now let $T'$ be a positive integer, and let
$c'_k(x)\in\{0,1\}$ be the coefficients in the binary expansion
\begin{equation}
\phi(x) = \sum_{k=1}^\infty c'_k(x) 2^{-k} \;.
\end{equation}
For $k=1,\ldots,T'$, we define unitary  
operations, $U_1,\ldots,U_{T'}$, on our quantum register by
\begin{equation} \label{eq:Uk}
U_k|x\rangle=e^{2\pi ic'_k(x)/2^k}|x\rangle\;.
\end{equation}
The operators $U_k$ are conditional phase shifts that can be realized as
quantum gate sequences using the classical algorithm to compute the phases 
$\phi(x)$ \cite{Cleve1998}.

The algorithm can now be described as follows.  Choose a suitable (small)
number, $a$, of {\em auxiliary qubits}, and define $L=\log_2N+a$.
Prepare a register of $L$ qubits in the state
\begin{equation}  \label{eq:Psi0}
  |\Psi^0\rangle=(2^aN)^{-1/2}\sum_{x=0}^{2^aN-1}|x\rangle \,.
\end{equation}
For $k=1,\ldots,T$, define the  Grover operator
\begin{equation}
\hat G(O_k,t_k)
  =\left((2|\Psi^0\rangle\langle\Psi^0|-\hat I)\hat O_k\right)^{t_k}\,,
\end{equation}
where $\hat I$ is the $L$-qubit identity operator, and where the integer
``times'' $t_k$ are defined in Eq.~(\ref{eq:t_k}) below. 
Apply the Grover operators successively to the register to create the state
\begin{equation} \label{eq:psiT}
|\Psi^T\rangle = \hat{G}(O_T,t_T)\cdots\hat{G}(O_1,t_1) |\Psi^0\rangle \;.
\end{equation}
Now measure the $a$ auxiliary qubits in the computational basis. If one of the
outcomes is 1 (the probability for this will be shown to be small), this stage
of the algorithm has failed, and one has to start again by preparing the
register in the state $|\Psi^0\rangle$ as in Eq.~(\ref{eq:Psi0}). Otherwise,
i.e., if all $a$ outcomes are 0, this stage of algorithm has succeeded, and
the resulting state of the remaining $L-a=\log_2N$ qubits, which we denote by
$|\tilde\Psi_{\rm r}\rangle$, will be a good approximation to the
real-amplitude state
\begin{equation}  \label{eq:PsiReal}
|\Psi_{\rm r}\rangle = \sum_{x=0}^{N-1} \sqrt{p(x)} \, |x\rangle \;,
\end{equation}
obtained from our target state $|\Psi\rangle$ by setting the phases $\phi(x)$
to zero.

The final stage of the algorithm adds phases to the state $|\tilde\Psi_{\rm
  r}\rangle$ by applying the operators $U_1,\ldots,U_{T'}$, 
\begin{equation}   \label{eq:addPhases}
|\tilde\Psi\rangle  =  U_1U_2\cdots U_{T'}|\tilde\Psi_{\rm r}\rangle \;.
\end{equation}

In the next section, we analyse the dependence of the state
$|\Psi^T\rangle$ on the numbers $t_k$, and thus motivate the
definition~(\ref{eq:t_k}). At the end of the section, we derive upper bounds
on the numbers $t_k$ and therefore on the required number of oracle calls.
In the final section, we derive a lower bound on the fidelity in terms of the
parameters $T$, $T'$ and $a$.

\section{Number of oracle calls}   \label{sec:t_k}

In the following we will use the notation $1:n$ to index an ordered sequence of
$n$ symbols, for example,
\begin{equation}
q_{1:n}=q_1,\dots,q_n\,.
\end{equation}
Using this notation, the statement $q_{1:n}=c_{1:n}$ means that $q_j=c_j$ for
any $j=1,\dots,n$.

We define a set of refined oracles, 
\begin{equation}
O_{q_{1:n}}(x)=\left\{\begin{array}{ll}
                       1& {\rm if\ } q_{1:n}=c_{1:n}(x) \,,\cr
                                           0& {\rm otherwise}\,,
                      \end{array}\right. 
\end{equation}
which can be expressed in terms of the oracles $O_k$ as follows.
\begin{equation}
O_{q_{1:n}}(x)=\prod_{k=1}^{n}|O_k(x)-1+q_k|\,.
\end{equation}
Let $\Omega_{q_{1:k}}$ be the set of values that are accepted by the
oracle $O_{q_{1:k}}$, i.e.
\begin{equation}
\Omega_{q_{1:k}}=\{x: O_{q_{1:k}}(x)=1\}\,.
\end{equation}
Furthermore, denote by $N^k_{q_{1:k}}$ the size of the set
$\Omega_{q_{1:k}}$,
\begin{equation}
N^{k}_{q_{1:k}}=\sum_{x}O_{q_{1:k}}(x)\,.
\end{equation}

The first stage of our algorithm takes the initial state $|\Psi^0\rangle$
through a series of intermediate states, $|\Psi^k\rangle$, to the state
$|\Psi^T\rangle$. Due to the properties of the Grover operators, $\hat
G(O_k,t_k)$, the intermediate states are of the form
\begin{equation}
|\Psi^k\rangle=
\sum_{q_{1:k}}\sum_{x\in\Omega_{q_{1:k}}}A^k_{q_{1:k}}\,|x\rangle\,,
\end{equation}
where
\begin{equation} \label{TargetAmplitudes}
A^k_{q_{1:k}}=B^k +\sum_{j=1}^{k} q_j h_j \,,
\end{equation}
where the {\em features\/} $h_j$ are positive numbers determined by the times
$t_k$, and where the $B^k$ are real numbers determined by the normalization
conditions $\langle\Psi^k|\Psi^k\rangle=1$.

We show next how the features $h_k$ depend on the numbers of Grover
iterations $t_k$. 

\subsection{General oracle}  \label{subsecBiham}

We will be using the following result of Biham {\it et.~al.}
\cite{Biham1999}. Consider an oracle $O$, which accepts $r$ values (out
of the total of $2^aN$, i.e., $\sum_{x=0}^{2^aN-1}O(x)=r$). We shall call such
values of $x$ {\em good}, as opposed to {\em bad\/} values of $x$ that are
rejected by the oracle. Using different notation for the coefficients of good
and bad states, we have that after $t$ Grover iterations an arbitrary quantum
state
\begin{equation} \label{initial}
|\Psi^{\rm ini}\rangle= \sum_{{\rm good\ }x} g^{\rm ini}_x|x\rangle
                           +\sum_{{\rm bad\ }x} b^{\rm ini}_x|x\rangle
\end{equation}
is transformed into
\begin{equation}
|\Psi^{\rm fin}\rangle=\hat{G}(O,t)|\Psi^{\rm ini}\rangle
=\sum_{{\rm good\ }x} g^{\rm fin}_x|x\rangle
                           +\sum_{{\rm bad\ }x} b^{\rm fin}_x|x\rangle.
\end{equation}
Let $\bar{g}^{\rm ini}$ and $\bar{b}^{\rm ini}$ be the averages of the initial
amplitudes of the good and the bad states respectively:
\begin{equation}
\bar{g}^{\rm ini}
=\frac{1}{r}\sum_{{\rm good\ }x}g^{\rm ini}_x\,,\hspace*{2cm}
\bar{b}^{\rm ini}=\frac{1}{2^aN-r}\sum_{{\rm bad\ }x}b^{\rm ini}_x\,,
\end{equation}
and similarly for the final amplitudes
\begin{equation}
\bar{g}^{\rm fin}
=\frac{1}{r}\sum_{{\rm good\ }x}g^{\rm fin}_x\,,\hspace*{2cm}
\bar{b}^{\rm fin}=\frac{1}{2^aN-r}\sum_{{\rm bad\ }x}b^{\rm fin}_x\,,
\end{equation}
Let us also define
\begin{equation}
\Delta g^{\rm ini}_x= 
g^{\rm ini}_x-\bar{g}^{\rm ini}\,,\hspace*{2cm}
\Delta b^{\rm ini}_x=b^{\rm ini}_x-\bar{b}^{\rm ini}\,.
\end{equation}
In other words, $\Delta g^{\rm ini}_x$ and $\Delta b^{\rm ini}_x$ define the
features of the initial amplitude functions $g^{\rm ini}_x$ and
$b^{\rm ini}_x$ relative to their averages $\bar{g}^{\rm ini}$ and
$\bar{b}^{\rm ini}$.  Biham {\it et.\ al.} have shown that the change of the
amplitudes is essentially determined by the change of the averages:
\begin{eqnarray} \label{BihamEquations}
g^{\rm fin}_x&=&\bar{g}^{\rm fin}+\Delta g^{\rm ini}_x \cr
b^{\rm fin}_x&=&\bar{b}^{\rm fin}+(-1)^t\Delta b^{\rm ini}_x\,,
\end{eqnarray}
where the averages $\bar{g}^{\rm fin}$ 
and $\bar{b}^{\rm fin}$ are given as follows. 
Define
\begin{eqnarray}
\omega&=&\arccos\left(1-\frac{2r}{2^{a}N}\right)\;,\\
\alpha
&=&\sqrt{|\bar{b}^{\rm ini}|^2+|\bar{g}^{\rm ini}|^2\frac{r}{2^aN-r}}\;,
   \label{eq:alpha}\\
\phi
&=&\arctan\left(\frac{\bar{g}^{\rm ini}}{\bar{b}^{\rm ini}}
    \sqrt{\frac{r}{2^aN-r}}\,\right) \label{phi} \;.
\end{eqnarray} 
The averages are given by
\begin{eqnarray} \label{FinalAverages}
\bar{g}^{\rm fin}&=&\sqrt{\frac{2^aN-r}{r}}\;\alpha\,\sin(\omega t+\phi)\,, \cr
\bar{b}^{\rm fin}&=&\alpha\,\cos(\omega t+\phi)\,.
\end{eqnarray}

These formulas allow us to calculate the number of Grover iterations $t$
from the ratios $\bar{g}^{\rm ini}/\bar{b}^{\rm ini}$ and $\bar{g}^{\rm
  fin}/\bar{b}^{\rm fin}$ as follows.  From Eqs.~(\ref{FinalAverages}) we have
\begin{equation}
\frac{\bar{g}^{\rm fin}}{\bar{b}^{\rm fin}}
=\sqrt{\frac{2^aN-r}{r}}\;\tan(\omega t+\phi)\,,
\end{equation}
which, together with~Eq.(\ref{phi}), gives
\begin{equation} \label{OmegaT}
\omega t=
\arctan\left(
\frac{\bar{g}^{\rm fin}}{\bar{b}^{\rm fin}}
\sqrt{\frac{r}{2^aN-r}}
 \right)
 -
 \arctan\left(\frac{\bar{g}^{\rm ini}}{\bar{b}^{\rm ini}}
    \sqrt{\frac{r}{2^aN-r}}\,\right)
 \,.
\end{equation}

\subsection{Formulas for $t_k$}

Consider the state $|\Psi^k\rangle$, i.e.\ the state that results after
building the first $k$ features using the oracles $O_1,\dots,O_k$.  Let
$\bar{g}^{\rm ini}_{k+1}$, $\bar{b}^{\rm ini}_{k+1}$ be the average amplitudes
of the ``good'' and ``bad'' states within $|\Psi^k\rangle$ with respect to the
oracle $O_{k+1}$.  By direct calculation we have
\begin{equation}
\bar{g}^{\rm ini}_{k+1}
=\frac{
\sum_{q_{1:k}}
A^k_{q_{1:k}}N^{k+1}_{q_{1:k}1}}{\sum_{q_{1:k}}
                                 N^{k+1}_{q_{1:k}1}}\;,\ \ \ \ \ \
\bar{b}^{\rm ini}_{k+1}
=\frac{
\sum_{q_{1:k}}
A^k_{q_{1:k}}N^{k+1}_{q_{1:k}0}}{\sum_{q_{1:k}}
                                 N^{k+1}_{q_{1:k}0}}\;,
\end{equation}
and therefore
\begin{equation}
\frac{\bar{g}^{\rm ini}_{k+1}}{\bar{b}^{\rm ini}_{k+1}}
=\frac{(N-N_{k+1})\sum_{q_{1:k}}
A^k_{q_{1:k}}N^{k+1}_{q_{1:k}1}}{
N_{k+1}
\sum_{q_{1:k}}
 A^k_{q_{1:k}}N^{k+1}_{q_{1:k}0}}\;.
\end{equation}
Similarly, in the case of the final averages
$\bar{g}^{\rm fin}_{k+1}$
and $\bar{b}^{\rm fin}_{k+1}$
we obtain
\begin{equation}
\frac{\bar{g}^{\rm fin}_{k+1}}{\bar{b}^{\rm fin}_{k+1}}
=\frac{(N-N_{k+1})\sum_{q_{1:k}}
A^{k+1}_{q_{1:k}1}N^{k+1}_{q_{1:k}1}}{
N_{k+1}
\sum_{q_{1:k}}
 A^{k+1}_{q_{1:k}0}N^{k+1}_{q_{1:k}0}}\;.
\end{equation}
Below we need expressions for the ratios ${\bar{g}^{\rm ini}_k}/{\bar{b}^{\rm
    ini}_k}$ and ${\bar{g}^{\rm fin}_k}/{\bar{b}^{\rm fin}_k}$, which follow
by substituting $k$ for $k+1$.  The number of Grover iterations,
$t_{k}$, required for converting the state $|\Psi^{k-1}\rangle$ into
$|\Psi^{k}\rangle$ can then be obtained from Eq.~(\ref{OmegaT}),
\begin{equation}
\omega_k t_k=
\arctan\left(
\frac{\bar{g}^{\rm fin}_k}{\bar{b}^{\rm fin}_k}
\sqrt{\frac{N_k}{2^aN-N_k}}
 \right)
 -
 \arctan\left(\frac{\bar{g}^{\rm ini}_k}{\bar{b}^{\rm ini}_k}
    \sqrt{\frac{N_k}{2^aN-N_k}}\,\right)
 \,,
\end{equation}
where
\begin{equation}
\omega_k=\arccos\left(1-\frac{2N_k}{2^{a}N}\right)\;.
\end{equation}

Of course these formulas for the integer times $t_k$ are useless by
themselves, because they depend on the coefficients $A^{k}_{q_{1:k}}$, which
are defined in terms of the unknown features $h_k$ [see
Eq.~(\ref{TargetAmplitudes})]. The following argument leads to an explicit
formula for the $t_k$.

By construction of the sets $\Omega_{q_{1:k}}$, the sums $\sum_{j=1}^{k} q_j
2^{-j}/\sqrt{\eta N}$ are $k$-bit approximations to the target amplitudes
$\sqrt{p(x)}$ for all $x\in\Omega_{q_{1:k}}$. We thus aim for the features
$h_j$ to be as close as possible to the values $2^{-j}/\sqrt{\eta N}$.  This
motivates the following choice for the $t_k$. Instead of the
amplitudes~(\ref{TargetAmplitudes}), we define
\begin{equation}
{A'}^k_{q_{1:k}}={B'}^k +\sum_{j=1}^{k} q_j2^{-j}/\sqrt{\eta N} \,,
\end{equation}
where the $h_j$ have been replaced by $2^{-j}/\sqrt{\eta N}$, and where 
the terms ${B'}^k$ are determined by the normalization conditions 
$\langle{\Psi'}^k|{\Psi'}^k\rangle=1$ for the states
\begin{equation}
|{\Psi'}^k\rangle=
\sum_{q_{1:k}}\sum_{x\in\Omega_{q_{1:k}}}{A'}^k_{q_{1:k}}\,|x\rangle\,.
\end{equation}
These states can be regarded as $k$-bit approximations to the intermediate
states $|\Psi^k\rangle$.
We thus get the following modified expressions for the average amplitudes.
\begin{equation}
\frac{\bar{g'}^{\rm ini}_{k+1}}{\bar{b'}^{\rm ini}_{k+1}}
=\frac{(N-N_{k+1})\sum_{q_{1:k}}
{A'}^k_{q_{1:k}}N^{k+1}_{q_{1:k}1}}{
N_{k+1}
\sum_{q_{1:k}}
{A'}^k_{q_{1:k}}N^{k+1}_{q_{1:k}0}}\;
\end{equation}
and
\begin{equation}
\frac{\bar{g'}^{\rm fin}_{k+1}}{\bar{b'}^{\rm fin}_{k+1}}
=\frac{(N-N_{k+1})\sum_{q_{1:k}}
{A'}^{k+1}_{q_{1:k}1}N^{k+1}_{q_{1:k}1}}{
N_{k+1}
\sum_{q_{1:k}}
{A'}^{k+1}_{q_{1:k}0}N^{k+1}_{q_{1:k}0}}\;.
\end{equation}
The final expression for the times $t_k$ is then
\begin{equation}   \label{eq:t_k}
t_k= \left\lfloor {1\over2} + {1\over\omega_k} \left(
\arctan\left(
\frac{\bar{g'}^{\rm fin}_k}{\bar{b'}^{\rm fin}_k}
\sqrt{\frac{N_k}{2^aN-N_k}}
 \right)
 -
 \arctan\left(\frac{\bar{g'}^{\rm ini}_k}{\bar{b'}^{\rm ini}_k}
    \sqrt{\frac{N_k}{2^aN-N_k}}\,\right)
\right) \right\rfloor
 \,,
\end{equation}
where the extra term $1/2$ combined with the $\lfloor\ldots\rfloor$ operation 
amounts to a rounding to the nearest integer.

The expressions for $t_k$ depend explicitly on the numbers $N^k_{q_{1:k}}$,
i.e., the numbers of points $x$ for which the $k$ most significant bits of
$\sqrt{\eta Np(x)}$ are given by $q_{1:k}$. If this {\em global information\/}
about the probabilities $p(x)$ is available, the version of our state
preparation algorithm described here will often be simpler than the original
version of the algorithm described in Ref.~\cite{SoklakovSchack}. If the
numbers $N^k_{q_{1:k}}$ are not available initially, they can be obtained via
the quantum counting algorithm \cite{Boyer1998}. In this case, the algorithm
described here loses much of its appeal. The analysis of the original
algorithm in Ref.~\cite{SoklakovSchack} includes bounds for the resources
required for the initial quantum counting step.

\subsection{Bound on the number of oracle calls}

We have, by definition,
\begin{equation}
\frac{2N_k}{2^aN}=1-\cos\omega_k=2\sin^2\frac{\omega_k}{2}\,.
\end{equation}
Since $x^2\geq\sin^2 x$ we obtain
\begin{equation} \label{Inequality112}
\omega_k \ge 2\sqrt{\frac{N_k}{2^aN}} \;.
\end{equation}
Furthermore, we have
\begin{equation}
\omega_k t_k \leq 2\pi\;,
\end{equation}
and hence
\begin{equation}    \label{eq:boundOnTk}
t_k\leq\frac{2\pi}{\omega_k}
\leq \pi\sqrt{\frac{2^aN}{N_k}}\,.
\end{equation}
The overall number of oracle calls is therefore bounded by the expression
$T'+T\pi\sqrt{{2^aN}/{N_k}}$.
A typical value for the fraction $N_k/N$ is $1/2$. The worst case for the
number of oracle calls corresponds to $N_k=1$, which is equivalent to Grover
database search \cite{Grover1997}. The efficiency of our algorithm can be
improved by ignoring very small values of $N_k$. Bounds for the corresponding
fidelity reduction have been derived in Ref.~\cite{SoklakovSchack}. An
analysis of the asymptotic number of oracle calls in the limit of large $N$ is
possible, e.g., for a sequence of states for which the parameter $\eta$ does
not depend on $N$ and the ratios $N_k/N$ tend to a constant $C_k$ as
$N\to\infty$. In this case, the fidelity bound~(\ref{eq:overallBound}) does
not depend on $N$. For the right-hand side of the bound~(\ref{eq:boundOnTk}),
we have then
\begin{equation}   
\pi\sqrt{\frac{2^aN}{N_k}} \to \pi\sqrt{2^a/C_k} \;\;\mbox{as}\;\;N\to\infty\;,
\end{equation}
i.e., the bound for the required number of oracle calls tends to a constant for
large $N$.

\section{Fidelity analysis}  \label{sec:fidelity}

In this section we derive a lower bound
for the fidelity $|\langle\tilde\Psi|\Psi\rangle|$.  
We start by considering the fidelity between the real-amplitude target state 
$|\Psi_{\rm r}\rangle$ defined in Eq.~(\ref{eq:PsiReal}) and the state 
$|\Psi^T\rangle$ resulting from the Grover iterations, but before the $a$
auxiliary qubits have been measured [see Eq.~(\ref{eq:psiT})]. It follows from
the discussion at the start of Sec.~\ref{sec:t_k} that 
$|\Psi^T\rangle$ can be written in the form
\begin{equation}             \label{eq:psiTalt}
|\Psi^T\rangle = \sum_{x=0}^{2^aN-1}
\Big(B^T+\sum_{j=1}^{T}c_j(x)h_j\Big)\,|x\rangle \;.
\end{equation}
The first step is done in 
subsection~\ref{sec:hDeltaBound}, where we show that the choice
Eq.~(\ref{eq:t_k}) 
for the integer times $t_k$ implies that the features $h_k$, for
$k=1,\ldots,T$, satisfy the inequalities
\begin{equation} \label{h_target_bound}
\left|h_k-\frac{2^{-k}}{\sqrt{\eta N}}\right|
<\frac{2^{1-a/2}}{\sqrt{\eta N}}\,.
\end{equation}
Subsection~\ref{sec:mainBound} uses this result to derive the fidelity bound
\begin{eqnarray}     \label{eq:firstBound}
|\langle\Psi_r|\Psi^T\rangle|
&=&\sum_{x=0}^{2^aN-1}\sqrt{p(x)}
\Big(B^T+\sum_{j=1}^{T}c_j(x)h_j\Big)\cr
&=&\sum_{x=0}^{N-1}\sqrt{p(x)}
\Big(B^T+\sum_{j=1}^{T}c_j(x)h_j\Big)\cr
&>& 1-3T\;\frac{2^{-a/2}}{\eta}\,,
\end{eqnarray}
where we have used the fact that $p(x)=0$ for $x\ge N$, and where we have
assumed that $T$ is chosen to be the smallest integer for which
\begin{equation}
\frac{2^{-T}}{2T^2}\leq 2^{-a}\,.
\end{equation}
One can of course use bigger values of $T$, but this would not improve the
performance of the algorithm as the fidelity of the state preparation
is limited by the choice of $a$ [see Eq.~(\ref{h_target_bound})].

The next step of the algorithm is the measurement of the auxiliary qubits. The
probability of failure, $p_{\rm fail}$, i.e.~the probability of obtaining a
nonzero result, is given by
\begin{equation}
p_{\rm fail}=(2^aN-N)\,|B^T|^2\;,
\end{equation}
where $B^T$ is the normalization term in the expression~(\ref{eq:psiTalt}) for
$|\Psi^T\rangle$.  Subsection~\ref{sec:mainBound} derives the following bound
on the failure probability.
\begin{equation}
p_{\rm fail} \leq 16T\;\frac{2^{-a/2}}{\eta}\;.
\end{equation}
If there is no failure, i.e., if the measurement outcome is zero, the
post-measurement state is given by 
\begin{equation}
|\tilde\Psi_{\rm r}\rangle = \frac{1}{\sqrt{1-p_{\rm fail}}}
\sum_{x=0}^{N-1} \Big(B^T+\sum_{j=1}^{T}c_j(x)h_j\Big) \,|x\rangle \;.
\end{equation}
Together with Eqs.~(\ref{eq:psiTalt}) and~(\ref{eq:firstBound}) it follows
directly that
\begin{equation}
|\langle\Psi_r|\tilde\Psi_r\rangle|  =  \frac{1}{\sqrt{1-p_{\rm
    fail}}}\;|\langle\Psi_r|\Psi^T\rangle|
\;>\; 1-3T\;\frac{2^{-a/2}}{\eta}\,.
\end{equation}
Finally, in subsection~\ref{sec:boundPhases} we combine this bound with a
simple analysis of the last stage in which the phases are added to the real
amplitudes of the state $|\tilde\Psi_r\rangle$. The result is the following
overall lower bound on the fidelity between the target state $|\Psi\rangle$
and the state $|\tilde\Psi\rangle$ prepared by the algorithm, 
\begin{equation}    \label{eq:overallBound}
|\langle\tilde\Psi|\Psi\rangle|  
>  \left(1-3T\;\frac{2^{-a/2}}{\eta}\right)( 1- 2^{-2T'-1} ) \;.
\end{equation}
This bound determines the performance of the state preparation algorithm
described in this paper.

\subsection{Upper bound on $\left|h_k-{2^{-k}}/{\sqrt{\eta N}}\right|$}
                                                \label{sec:hDeltaBound}

Consider the development of a single feature, $h$, in $t$ Grover iterations
based on an oracle~$O$. Let $r$ be the number of good states, or solutions, of
$O$. It follows from Eqs.~(\ref{FinalAverages}) that $h$ depends on $t$ via
\begin{equation}
h(t)=\alpha \sqrt{2^aN/r}\;\sin(\omega t-\xi)\,,
\end{equation}
where the values of $\alpha$ and $\xi$ depend on the initial average
amplitudes $\bar{g}^{\rm ini}$ and $\bar{b}^{\rm ini}$ of the good and the bad
states with respect to the oracle $O$. According to Eq.~(\ref{eq:alpha}) we
have
\begin{equation}   \label{eq:alphaSquared}
\alpha^2=|\bar{b}^{\rm ini}|^2+
|\bar{g}^{\rm ini}|^2\frac{r/N}{2^a-r/N}\,.
\end{equation}
The average amplitude of ``bad'' states is bounded as
\begin{equation}
\bar{b}^{\rm ini}\le\frac{1}{\sqrt{2^aN}}\,,
\end{equation}
and the maximum possible value of the average amplitude of ``good'' states is
bounded as
\begin{equation}
\bar{g}^{\rm ini}\le\frac{1}{\sqrt{\eta N}}\,.
\end{equation}
Hence Eq.~(\ref{eq:alphaSquared}) implies
\begin{eqnarray}   \label{bound_on_alpha}
\alpha^2&\leq&\frac{1}{2^aN}+\frac{r/N}{\eta N(2^a-r/N)}\cr
&\leq&\frac{1}{2^aN}+\frac{1}{\eta N(2^a-1)}\cr
&\leq&\frac{1}{2^aN}+\frac{2}{\eta 2^a N}\cr
&<&\frac{4}{\eta 2^aN} \;.
\end{eqnarray}
Since in our algorithm the value of $t_k$ is rounded to the nearest integer,
$h_k=h(t_k)$ will rarely coincide with the target value of $2^{-k}/\sqrt{\eta
  N}$.  The  mistake, however, can be bounded as
\begin{equation}
\left|h_k-\frac{2^{-k}}{\sqrt{\eta N}}\right|
\leq \max |h(t+1)-h(t)|\,,
\end{equation}
where the maximum is taken with respect to the quantities $\alpha$, $\omega$,
$\xi$, $r$ and $t$. The parameters characterizing the algorithm, $a$, $\eta$
and $N$, are being kept constant. Using~(\ref{bound_on_alpha}) we
obtain
\begin{eqnarray}
\left|h_k-\frac{2^{-k}}{\sqrt{\eta N}}\right|
&<&\max\frac{2}{\sqrt{\eta r}}\Big{|} \sin\big((\omega t-\xi)+\omega\big)
-\sin(\omega t-\xi)
                                  \Big{|}\cr
&\leq&\max\frac{2}{\sqrt{\eta r}}\sin(\omega/2)\,,
\end{eqnarray}
where the last inequality follows from the properties
of the sin function.
Since $0\le\omega\le\pi$ we have 
\begin{equation}
\sin(\omega/2)=\sqrt{\frac{1-\cos\omega}{2}}=\sqrt{\frac{r}{2^a N}}\,,
\end{equation}
which implies the bound
\begin{equation}     \label{eq:boundSameAgain}
\left|h_k-\frac{2^{-k}}{\sqrt{\eta N}}\right|
<\frac{2^{1-a/2}}{\sqrt{\eta N}}
\end{equation}
as required. 

\subsection{Lower bound on $|\langle\Psi_r|\Psi^T\rangle|$}
                                                  \label{sec:mainBound}
Directly from the definitions we have
\begin{eqnarray}
\langle\Psi_r|\Psi^T\rangle
&=&\sum_{x=0}^{2^aN-1}\sqrt{p(x)}
\Big(B^T+\sum_{j=1}^{T}c_j(x)h_j\Big)\cr
&=&\sum_{x=0}^{N-1}\sqrt{p(x)}
\left(B^T+\sum_{j=1}^{T}c_j(x)\frac{2^{-j}}{\sqrt{\eta N}}
+
\sum_{j=1}^{T}c_j(x)\left(h_j-\frac{2^{-j}}{\sqrt{\eta N}}\right)\right)\,,
\end{eqnarray}
where we have used the fact that $p(x)=0$ for $x\geq N$.
Let us define
\begin{equation} \label{b_and_delta}
b(x)=\sum_{j=T+1}^{\infty}c_j(x)\frac{2^{-j}}{\sqrt{\eta N}}\;,\ \ \ \ \
\delta(x)=\sum_{j=1}^{T}c_j(x)\left(h_j-\frac{2^{-j}}{\sqrt{\eta N}}\right)\,.
\end{equation}
Since
\begin{equation}
\sum_{j=1}^{T}c_j(x)\frac{2^{-j}}{\sqrt{\eta N}}
=\sqrt{p(x)}-b(x)\,,
\end{equation}
and since $\sum_{x=0}^{N-1}p(x)=1$ we have
\begin{equation}
|\langle\Psi_r|\Psi^T\rangle|
\geq 1-
\left|\sum_{x=0}^{N-1}\sqrt{p(x)}
\Big(B^T - b(x)\Big)\right|
-
\sum_{x=0}^{N-1}\sqrt{p(x)}\,
\left|\delta(x)\right|\,.
\end{equation}
Using the bound~(\ref{eq:boundSameAgain}), one can show
that
\begin{equation} \label{bound_on_delta}
|\delta(x)|\leq 2T\frac{2^{-a/2}}{\sqrt{\eta N}}\,,
\end{equation}
and therefore, using $\sqrt{p(x)}\le1/\sqrt{\eta N}$, 
\begin{equation} \label{Psi_r_PsiT_almost}
|\langle\Psi_r|\Psi^T\rangle|
\geq 1-
\left|\sum_{x=0}^{N-1}\sqrt{p(x)}
\Big(B^T - b(x)\Big)\right|
-
2T\frac{2^{-a/2}}{\eta}\,.
\end{equation}
The function $b(x)$ satisfies the bounds
\begin{equation}    \label{bound_on_b}
0\leq b(x)\leq 2^{-T}/\sqrt{\eta N}\,.
\end{equation}
In order to find the lower bound on $|\langle\Psi_r|\Psi^T\rangle|$ {from}
Eq.~(\ref{Psi_r_PsiT_almost}) we need to calculate~$|B^T|$. This can be done
by examining the normalization condition $\langle\Psi^T|\Psi^T\rangle=1$ which
reads
\begin{equation}
\sum_{x=0}^{2^aN-1}
\left(
B^T+\sum_{j=1}^Tc_j(x)h_j
\right)^2=1\,.
\end{equation}
Using the definitions~(\ref{b_and_delta}), this can
be rewritten as
\begin{equation}
\sum_{x=0}^{2^aN-1}
\left(
B^T+\sqrt{p(x)}-b(x)+\delta(x)
\right)^2=1\,.
\end{equation}
This leads to a quadratic equation for $B^T$:
\begin{equation}
(B^T)^2+2UB^T+V=0\,,
\end{equation}
where
\begin{eqnarray}
U&=&\frac{1}{2^aN}\sum_{x=0}^{N-1}\Big(\sqrt{p(x)}+\delta(x)-b(x)\Big)\,,\\
V&=&\frac{1}{2^aN}\sum_{x=0}^{N-1}
\Big(
2\sqrt{p(x)}+\delta(x)-b(x)\Big)\Big(\delta(x)-b(x)\Big)\,.
\end{eqnarray}
Since $\sqrt{p(x)}-b(x)\geq 0$, using the
inequalities~(\ref{bound_on_delta}) and~(\ref{bound_on_b})
together with the bound $\sqrt{p(x)}\leq 1/\sqrt{\eta N}$
we obtain
\begin{eqnarray}
-2T\frac{2^{-3a/2}}{\sqrt{\eta N}} \;\;\leq 
&U&
\leq\;\;\frac{2^{-a}}{\sqrt{\eta N}}
(1+2T\,2^{-a/2})\,,\\
-4T\frac{2^{-a}}{\eta N}
\Big(\frac{2^{-T}}{2T}+2^{-a/2}\Big)\;\;\leq
&V&
\leq\;\; 4T\frac{2^{-3a/2}}{\eta N}(1+T\,2^{-a/2})\,.
\end{eqnarray}
As mentioned earlier, we assume that $T$ is chosen to be the smallest
integer for which
\begin{equation}
\frac{2^{-T}}{2T^2}\leq 2^{-a}\,.
\end{equation}
The above bounds can then be simplified as follows.
\begin{eqnarray}
|U|&\leq&2\,\frac{2^{-a}}{\sqrt{\eta N}}\;,\\
|V|&\leq&8T\,\frac{2^{-3a/2}}{\eta N}\;.
\end{eqnarray}
The value of $B^T$ therefore satisfies
the bound
\begin{eqnarray}  \label{bound_on_BT}
|B^T|&\leq& |U|+\sqrt{U^2+|V|}\cr
     &\leq& 2\,\frac{2^{-a}}{\sqrt{\eta N}}
          +\sqrt{9T\,2^{-3a/2}/(\eta N)}\cr
         &\leq& 4\sqrt{T}\;\frac{2^{-3a/4}}{\sqrt{\eta N}}\;.
\end{eqnarray} 
Using this bound together with~(\ref{bound_on_b})
we obtain from Eq.~(\ref{Psi_r_PsiT_almost}) the result
\begin{eqnarray}
|\langle\Psi_r|\Psi^T\rangle|
&\geq& 1-
\frac{1}{\eta}\left(
4\sqrt{T}2^{-3a/4}+2^{-T}
+
2T2^{-a/2}\right)\cr
&>& 1-3T\;\frac{2^{-a/2}}{\eta}\,.
\end{eqnarray}

Directly {from} Eq.~(\ref{bound_on_BT})
we obtain the upper bound on the failure probability, 
\begin{equation}
p_{\rm fail}=(2^aN-N)|B^T|^2\leq 16T\;\frac{2^{-a/2}}{\eta}\;.
\end{equation}

\subsection{Adding phases}   \label{sec:boundPhases}

The state $|\tilde\Psi_r\rangle$ resulting from the measurement of the $a$
auxiliary qubits has real amplitudes, i.e., it is of the form
\begin{equation}
|\tilde\Psi_r\rangle
= \sum_x \sqrt{\tilde p(x)} \,|x\rangle  \;.
\end{equation}
The final stage of the algorithm, see Eq.~(\ref{eq:addPhases}), turns 
$|\tilde\Psi_r\rangle$ into the final state $|\tilde\Psi\rangle$, which can 
be written as
\begin{equation}
|\tilde\Psi\rangle
= \sum_x \sqrt{\tilde p(x)}\,\exp[2\pi i\tilde \phi(x)] \,|x\rangle  \;,
\end{equation}
where the $\tilde\phi(x)$ are $T'$-bit approximations to the target phases
$\phi(x)$, i.e.,
\begin{equation}
|\phi(x)-\tilde\phi(x)| \le 2^{-T'} \;.
\end{equation}
Putting everything together, we find
\begin{eqnarray}
|\langle\tilde\Psi|\Psi\rangle| 
&=& \Big| \sum_x \sqrt{p(x)\tilde p(x)} \,
           \exp[2\pi i(\phi(x)-\tilde \phi(x))] \Big| \cr
&\ge& \sum_x \sqrt{p(x)\tilde p(x)} \, 
      \cos[\phi(x)-\tilde \phi(x)]  \cr
&\ge& \sum_x \sqrt{p(x)\tilde p(x)} \, 
      \big( 1- [\phi(x)-\tilde \phi(x)]^2/2 \big) \cr
&\ge& \sum_x \sqrt{p(x)\tilde p(x)} \, 
      ( 1- 2^{-2T'-1} ) \cr
&=& |\langle\Psi_r|\tilde\Psi_r\rangle| \; ( 1- 2^{-2T'-1} ) \cr
&>& \left(1-3T\;\frac{2^{-a/2}}{\eta}\right)( 1- 2^{-2T'-1} ) \;,
\end{eqnarray}
which is the required lower bound for the overall fidelity of the prepared
state. 

%\bibliographystyle{prsty}
%\bibliography{/home/rschack/lit/p}

\end{document}